# Near-field imaging of optical nano-cavities in Hyperuniform disordered materials

N. Granchi[1*], M. Lodde[2], K. Stokkereit[3], R. Spalding[3], P.J. van Veldhoven[2], R. Sapienza[4], A. Fiore[2], M. Gurioli[1], M. Florescu[3] and F. Intonti[1]

[1]Department of Physics and Astronomy and LENS, University of Florence, Sesto Fiorentino (FI), Italy
[2]Department of Applied Physics and Science Education, Eindhoven Hendrik Casimir Institute, Eindhoven University of Technology, Eindhoven, NL
[3]Advanced Technology Institute and Department of Physics, University of Surrey, Surrey, UK
[4]The Blackett Laboratory, Department of Physics, Imperial College London, UK
*granchi@lens.unifi.it

Abstract: Hyperuniform disordered photonic materials have recently been shown to display large, complete photonic bandgaps and isotropic optical properties, and are emerging as strong candidates for a plethora of optoelectronic applications, making them competitive with many of their periodic and quasiperiodic counterparts. In this work, high quality factor optical cavities in hyperuniform disordered architectures are fabricated through semiconductor slabs and experimentally addressed by Scanning Near-field Optical Microscopy. The wide range of confined cavity modes that we detect, arise from carefully designed local modifications of the dielectric structure. Previous works on hyperuniform disordered photonic systems has previously identified several Anderson localized states spectrally located at the PBG edges with relatively high quality factors. In this work, by engineering the structural parameters of the cavity, we achieve an experimental quality factor of order 6000 (higher than the one of the Anderson states) and we demonstrate that three types of localized modes of different nature coexist within a small area and in a relatively narrow spectral window of the disordered correlated system. Their compatibility with general boundary constraints, in contrast with ordered architectures that suffer strict layout constraints imposed by photonic crystal's axes orientation, makes optical cavities in disordered hyperuniform patterns a flexible optical insulator platform for planar optical circuits.

INTRODUCTION.
Photonic crystal cavities (PCCs) are state-of-the-art devices to strongly localize electromagnetic fields in volumes below a cubic optical wavelength of light, acting as efficient nano-resonators with high Q factors ($Q_f \approx 10^6$). Thanks to the photonic band gap (PBG) confinement effect, the well-controlled photon mode profiles in PCCs are extremely important for quantum electro-dynamics (QED) devices and integrated photonic applications, from lasing to optical fibers [1-5]. However, the anisotropy of PBGs caused by the lower-order rotational symmetry in wave-vector k-space has often hampered practical implementations, driving many efforts in developing alternative platforms like quasiperiodic photonic crystals [5-9] and correlated disordered photonic media [10]. Indeed, applications that demand cladding of some different heterostructures (like different types of photonic crystals or quasicrystals) require very isotropic band gaps to engineer point and line defects that span over larger spectral ranges and that are more robust against imperfections [11-12].
One approach of achieving more isotropic PBGs is to increase the rotational symmetry of the underlying structure as in the case of quasicrystals, which maintain the strong scattering of light while increasing the long-range orientational order in the system [13] and provide more efficient and uniform in-plane confinement in all directions, which could be beneficial for achieving lasing properties of lower threshold and higher quality factors [14,15]. A different approach has emerged in the last years, which is to erase the precise rotation symmetry while still preserving the PBG by using a special class of disordered photonic materials called Hyperuniform Disordered (HuD) systems [16]. These systems located in-between random structures and perfectly ordered photonic crystals, have emerged as a valid alternative solution to the problem: in particular, the structures built around *stealthy* HuD systems, whose name relates to their property of being transparent to incident radiation for a certain range of wavevectors k, have recently been shown to display large isotropic band gaps comparable in width to band gaps found in photonic crystals [10, 16-18]. In the context of optical confinement applications, HuDs provide two main advantages : (i) the large band gaps found in these structures are facilitated by the hyperuniform geometrical properties of the underlying point-pattern template upon which the structures are built [11,19] and (ii) the intrinsic geometrical statistical isotropy of the stealthy HuD point patterns induces a consequent statistical isotropy of the photonic properties of the dielectric structures, and this is highly relevant for a series of novel photonic functionalities, as their compatibility with general boundary constraints can provide a flexible optical insulator platform for planar optical circuits [20]. Many experimental works have focused on light confinement in 3D and 2D correlated disorder photonic systems [21-25]. Another unique feature exhibited by HuD systems and recently demonstrated in luminescent HuD photonic networks is that they support modes more robust against local perturbations and fabrication induced disorder with respect to their disordered and ordered counterparts [20]; eventual flaws that could seriously degrade the optical characteristics of photonic crystal devices are likely to have less effect on disordered hyperuniform structures, therefore relaxing fabrication constraints.

Consequently, the concept of optical cavities in HUD photonic materials proposed in Refs. [17,19] may offer a novel route that could address many of the roadblocks in the field of optical cavities and could provide many opportunities yet to be explored experimentally. In this work, we address this task by designing and fabricating optical cavities in HuD photonic networks on slab technology [26, 27], and explore their properties by performing a Scanning Near-field Optical Microscopy (SNOM) study on cavities with different structural parameters in order to obtain accurate knowledge of the design configuration that can maximize their quality factor. The wide range of confined cavity modes detected here and previously identified only theoretically [19], is obtained through carefully designed local modifications of the HuD dielectric structure and accurate fabrication and experimental characterization of the localized mode properties. So far, the highest experimental value of $Q_f$ obtained for not engineered localized modes in HuD systems on slab has been detected as 1500 [26]. Thanks to this analysis, we are able to engineer the structural parameters that allow to achieve an experimental maximum value of $Q_f \approx 6000$.

THEORY AND DESIGN.

The structures investigated comprise planar HuD dielectric networks designed through a tessellation protocol described in [10]. Analogously to the lattice constant in photonic crystals, we define a length scale $a = L/\sqrt{N}$, such that an *N*-point hyperuniform pattern in a square box of side length *L* has a scatterer density of $1/a^2$. The point pattern employed here contains *N* = 500 points and belongs to the sub-category of HuD systems known as "stealthy" [28, 29]. For stealthy hyperuniform patterns, the structure factor $S(k)$ is statistically isotropic, continuous and precisely equal to zero for a finite range of wavenumbers smaller than a certain critical wavevector $k_c$, i.e., $S(k < k_c) = 0$. The stealthiness parameter $\chi$ is defined as the ratio between the number of k-vectors for which the structure factor $S(k)$ is constrained to vanish and the total number of k-vectors. Analogously to the case photonic crystals, the photonic modal resonances supported by HuD structures can be tuned by changing two structural parameters [19, 26, 30]: the width of the decorating dielectric walls in the network, i.e. $w$, and a global scaling factor $GF$, which multiplies the size *L* and yields the final size of the HuD domain. We start with the 2D band structure analysis of the unperturbed HuD network with $\chi = 0.5$ and $a$ = 380nm, shown in Fig.1a. The walls are decorated with dielectric material of index of refraction n=3.4, and their width is $w = 0.36a$. In Fig.1b, we present the corresponding 2D photonic band structure, calculated using the planewave expansion software MPB [30]. The band structure calculations are done using a supercell approach, which results in a folding of the bands. However, the photonic band gap (white region) is not affected by the folding and is located between bands (red lines) *N* and *N+1*, where *N* is the number of scattering cells in the supercell (here *N*=500). Fig. 1c shows the cavity design, i.e. the HuD network of Fig.1a in which we placed an engineered defect at its center. The optical cavity is created firstly by filling one of the air domains, and after placing an inner circular air hole (of radius $R$) at its center. Then we employ a methodology for achieving optimal designs similar to the conventional one used for photonic crystal cavities [1] of smoothing the transition between the cavity and the HuD surround to minimize the leakage of the confined mode in the vertical direction. This involves slightly reducing the size of the adjacent holes and shifting their position outwards along the lattice directions. In this disordered case there are no lattice directions, and we instead shift the six shrunken cells, surrounding the defect, along the vector given by the center of mass of the cavity to the center of mass of the neighboring cell. The cells are shrunken to 52% of the size of a cell with infinitesimal walls and are shifted by 8% of the length of the vector outwards [19]. The central defect is designed such that it supports several modes, as predicted by the 2D bandstructure shown in Fig. 1d, where several bands appear inside the PBG. We focus on the first four modes, that we label accordingly with the 2D profiles of the magnetic field component $H_z$ (Fig.1e): dipole-like (D), hexapole-like (H), quadrupole-like (Q) and octupole-like (O) [6]. Interestingly, similar elementary localized modes have already been detected in photonic quasicrystal single cells [14], proving the main advantages related to the insensitivity to the propagation direction and to the long coherent-interaction-range order with respect to their periodic counterparts.

3D samples are obtained by adapting the 2D theoretical designs, and extruding them in the vertical direction to produce a slab of finite thickness employing a process described in ref. [26]. The effects of the finite height of the slab on the spectral position of the cavity's modes can be compensated by tuning an additional overall scaling factor $GF$; is a multiplying factor to the original size of the unitary cell $L$ that as a result increases ($GF > 1$) or decreases ($GF < 1$) the final dimension of the cell. This determines a rigid shift of all the bandstructure in terms of PBG. Hence, In order to finely tune the cavity resonances and to find the best conditions that maximize the experimental $Q_f$ we change the parameters $GF$, $R$ and $w$ (the last two parameters are highlighted in the upper right inset of Fig.1c). Consequently, we design several types of cavities with three different values of $w$ ($0.36a$, $0.4a$ and $0.44a$) and three different values of $R$ ($0.2a$, $0.3a$ and $0.4a$) that span over values of $GF$ from 0.9 to 1.1.

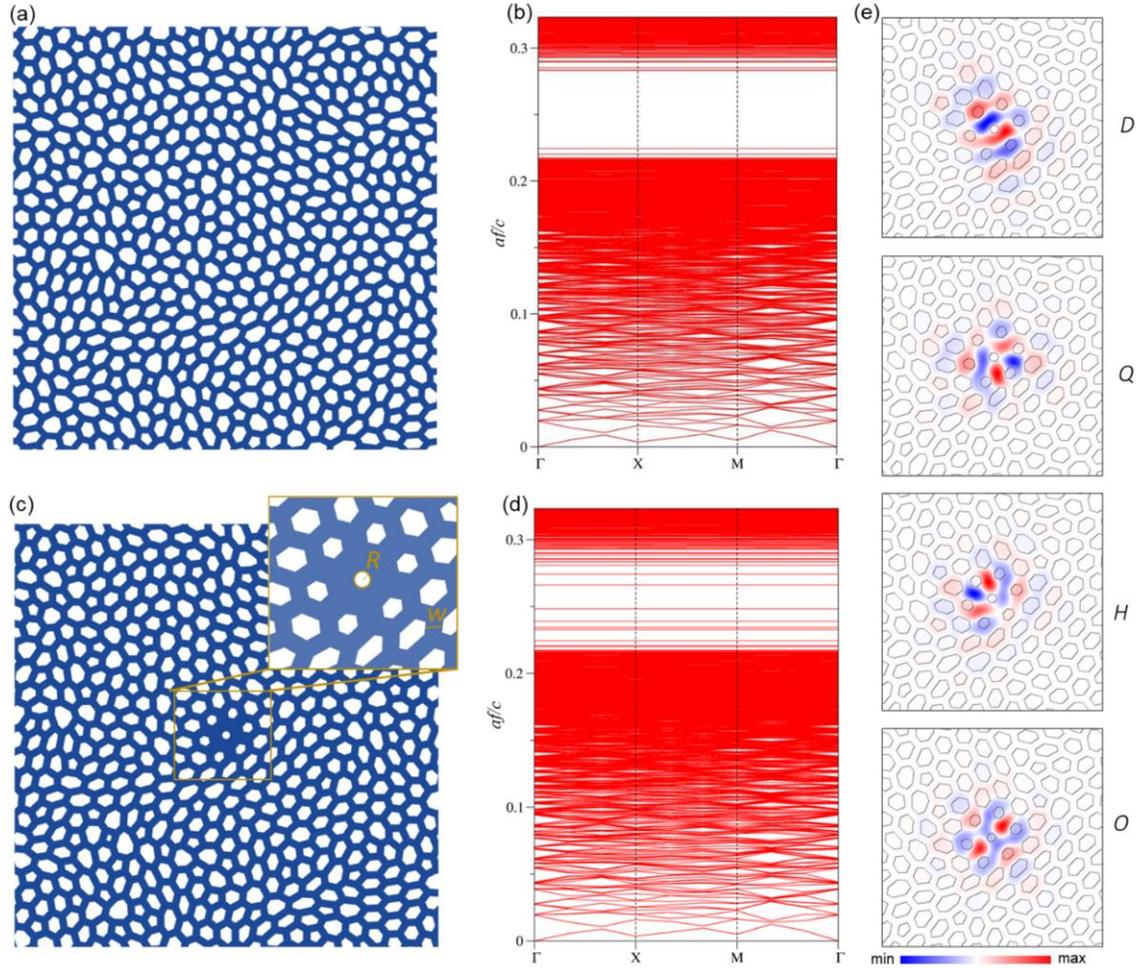

Figure 1: (a) Sketch of the 2D theoretical design of the hyperuniform disordered network, with length scale $a$ = 380nm, wall thickness $w = 0.36a$ and stealthiness $\chi = 0.5$. (b) 2D Bandstructure corresponding to the design on the left. (c) Sketch of the 2D theoretical design for an optical cavity embedded in the HuD surround. In the inset are highlighted the radius of the central hole $R = 0.2a$ and the thickness of the dielectric walls $w$. (d) 2D Bandstructure (MPB) corresponding to the design on the left. (e) Results of 2D simulations: Magnetic field $H_z$ (component normal to the structure plane) of the bands located inside the PBG corresponding to the four cavity modes. The modes are labelled, from the top to the bottom, in order of increasing frequency, D (dipole-like), Q (quadrupole like), H (hexapole-like) and O (octupole-like).

EXPERIMENT AND DISCUSSION

The structures fabricated consist of GaAs-based heterostructures: high-density InAs QDs emitting at 1300 nm are grown by molecular beam epitaxy at the center of a 200-nm-thick GaAs membrane and cover the entire area of the samples. The HuD structures with optical cavities under consideration are patterned with electron-beam lithography, reactive-ion etching and subsequent selective etching of a 3 μm-thick AlGaAs sacrificial layer [27]. Fig.2a shows a scanning-electron-microscopy (SEM) top view image of one of the fabricated samples, with nominal lattice constant $a$ = 380 nm, $GF$=0.9, wall thickness $w = 0.36a$ and radius of the central hole $R = 0.2a$ (same parameters as the design shown in Fig.1). A room temperature commercial SNOM (*Twinsnom*, OMICRON) is used in an illumination-collection geometry (see sketch of Fig.2b). The sample is excited with light from a diode laser (780 nm) coupled into a chemically etched optical fiber, which allows a direct measurement of the local density of states (LDOS) through the spectral shift maps [31,32]. It has been demonstrated that the tip-induced spectral shift is an excellent tool to measure the local electric field intensity, as the strength of the tip-induced spectral shift is proportional to the electric field intensity of the eigenmode itself. In our experiment, photoluminescence (PL) spectra from the sample are collected at each tip position through the same probe and the PL signal dispersed by a spectrometer is detected by a liquid nitrogen cooled InGaAs array with a high spectral resolution of 0.1 nm. This allow to reconstruct PL maps, from which we can extract the spectral shift maps to achieve a reliable imaging of the LDOS [31-,32].

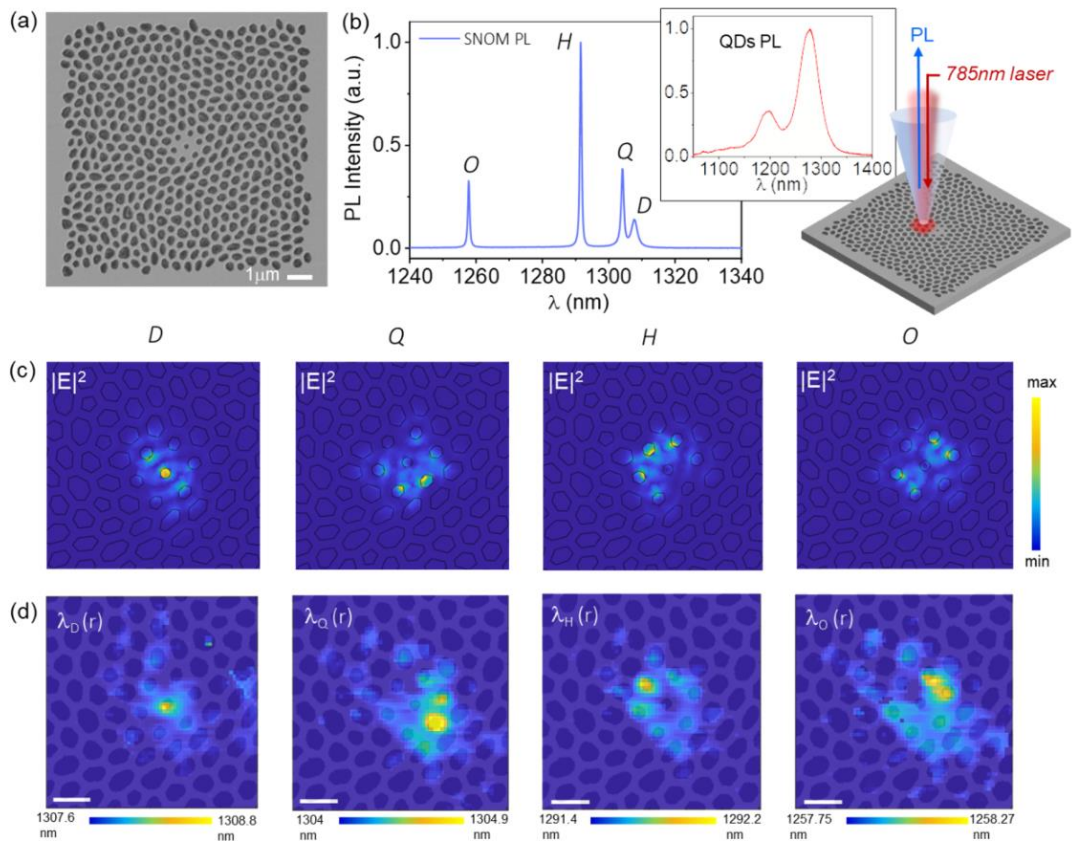

Figure 2: (a) SEM top view image of one of the fabricated samples, with nominal lattice constant $a$ = 380 nm, wall thickness $w = 0.36a$, radius of the central hole $R = 0.2a$ and $GF = 0.95$. (b) Typical SNOM PL spectrum acquired on top of the cavity, showing the four main resonances. The upper left inset shows the PL spectrum of the quantum dots used, on top of a sketch of the SNOM tip used for illumination/collection. (c) 3D FEM simulated maps of the electric field intensity distribution for the four resonances. (d) SNOM spectral shift maps revealing the LDOS of the four resonances. All scale bars correspond to 1 $\mu m$.

In the inset graph of Fig.2b it is shown the PL emission spectrum of the QDs collected in an unpatterned area of the sample. In Fig.2b we report a typical PL spectrum acquired on top of the cavity, where four sharp peaks are detected. The morphological information given by the SNOM topography, combined with the SNOM optical maps, allows us to establish that the four resonances correspond to the four cavity modes predicted our theoretical investigation: the dipole-like mode D ($\lambda$ = 1307.4 nm), the quadrupole-like mode Q ($\lambda$ = 1304.5 nm), the hexapole-like mode H ($\lambda$ = 1291.7 nm) and the octupole-like mode O ($\lambda$ = 1258.2 nm).We note that these resonances are located in the PBG spectral region (see Fig.1a). For each peak of the PL spectrum, we performed a single Lorentzian fit. This allows to reconstruct the spectral shift maps induced by the SNOM tip (overlapped in transparency with the SEM image of the sample and reported in Fig.2c) and achieve accurate knowledge of the LDOS spatial distribution of every resonance of the cavity [31,32]. With this method, from the Lorentzian fit of each peak we can calculate the $Q_f$; notably, the four sharp peaks of Fig.2b exhibit a $Q_f$ of 700, 1500, 2500 and 2150 (in order of increasing energy). We performed Finite Element Method (FEM) simulations of the GaAs slab patterned with the nominal design of Fig.1b. By comparing the experimentally inferred spectral shift maps to the FEM simulated maps of the electric field intensity distribution (reported in Fig.2d), we can conclude that the sub-micrometric details of the design are faithfully reproduced. Our investigation reveals not only the exact location each mode's hotspot, but also that the symmetry of Q and H modes is not entirely complete as their spatial distribution is unbalanced towards opposite directions instead of being distributed over the entire defect (an consequence of the intrinsic disorder in the HuD structure also evident in the simulated maps).

Before exploring the quality factors of HuD optical cavities, we analyze the light localization mechanisms. In photonic crystals, light with frequencies above or below the band edges are propagating modes that are transmitted through them, but they can become localized in the case of 2D hyperuniform disordered structures [26]. Here, we take a step further by introducing a cavity in the same HuD environment. In Fig. 3a we plot the Inverse Participation Ratio ($IPR$) as a measure of light localization [33-35, 26] for all the modes of the HuD network slab (same structures as Fig.1 and 2) obtained with FEM simulations. Values of $IPR$ close to one indicate the electric field intensity is nearly constant throughout the system volume, while higher values are associated with the mode profiles confined in a small volume. The blue dots indicate modes above and below the PBG; the expected trend portrays dielectric (air) modes with high $IPR$ ($IPR > 50$, i.e., strongly localized modes) at the upper (lower) PBG edge. The value of the $IPR$ decrease and reach the value of $\approx 1$ as the frequency increases and the optical modes enter a diffusive regime and become delocalized. The red dots (reported on a different scale) indicate cavity modes and the accidental mode localized over a topological defect (T) [10,26].

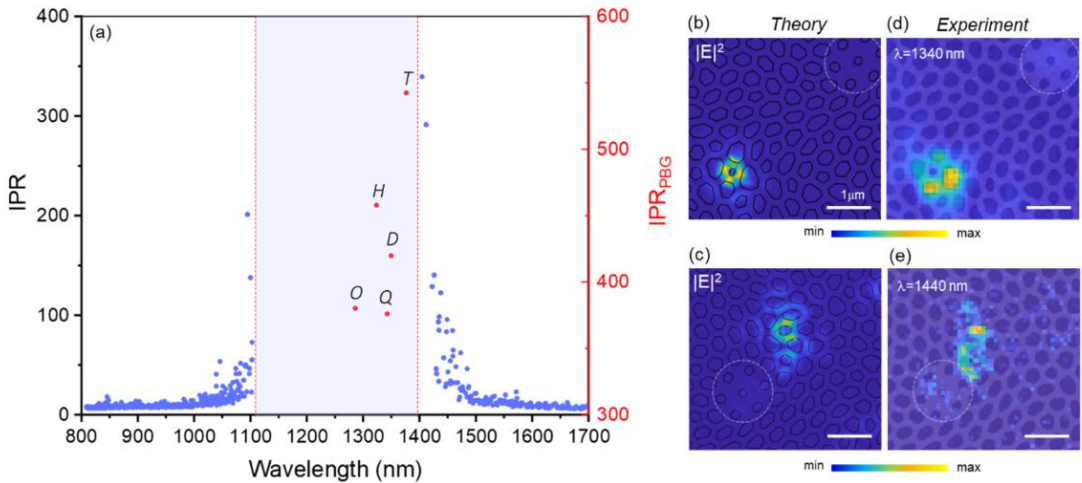

Figure 3: (a) Inverse Participation Ratio ($IPR$) obtained with 3D FEM simulations for every mode of HuD system (same parameters as Fig.1 and 2) (blue dots), for the cavity modes (O, H, Q and D) and the accidental mode T located over the topological defect (red dots). The light purple rectangle highlights the PBG. (b) 3D FEM maps of the electric field intensity of mode T and of one of the dielectric Anderson modes at the PBG upper edge, respectively first and second row. In the maps, the presence of the cavity is highlighted with a dashed white circle. (c) SNOM PL maps of mode T and of the Anderson dielectric mode, filtered around the central wavelengths of the corresponding peaks, $\lambda$ = 1304 nm and 1347 nm.

Clearly, these strongly localized modes ($IPR \approx 400\text{-}550$) coexists with all the other modes supported by the HuD network. This remarkable behavior is further explored in FEM simulations: in Fig. 3b and 3c we respectively present FEM maps of the electric field intensity of the T mode and of one of the dielectric Anderson modes at the PBG upper edge; the location of the cavity is highlighted with a dashed white circle. We compare the FEM maps with the SNOM PL maps filtered around the central wavelengths of each mode, 1304 nm for the T mode (Fig.3d) and 1347 nm for the dielectric Anderson mode (Fig.3e). As shown in Ref. [26], the T mode, that is pushed inside the PBG for patterns with $\chi = 0.5$, displays the tightest localization, as confirmed by the graph of Fig. 3a. In the experiment this mode is spectrally found between the cavity modes, and not at longer wavelengths as theory predicts; this is due to the extreme sensitivity of the accidental modes to disorder (in this case, induced by fabrication). Fig.3e shows the SNOM PL map of the Anderson mode. The map of this mode, localized in the dielectric part of the network and spatially located near the cavity, is definitely noisy: this is because of the low PL signal collected in a spectral region at wavelengths longer than the QDs PL first band. Nevertheless, the results shown in Fig. 2 and 3 clearly demonstrate the ability of the HuD structures to support a variety of localized modes: three types of modes coexist in a small area and in a relatively narrow spectral window: light is confined over a topological defect (T) of the network, undergoes Anderson localization, and, finally, can be trapped with a high $Q_f$ in an engineered cavity.

We now address the dependence of the cavity modes spectral positions on the variation of the two structural parameters previously introduced, the wall thickness $w$ and the central hole radius $R$. We first analyze the spectral positions of each of the resonances and their behavior. Here, we focus on structures with a fixed value of $GF = 1$, since, as argued before, the global scaling factor, $GF$, determines the spectral positioning of the modes that achieve maximum $Q_f$ values. We performed FEM simulations of nine cavities with three different values of $w$ and three different values of $R$, specifically: $w$=0.36$a$, 0.4$a$ and 0.44$a$, and $R$=0.2$a$, 0.3$a$ and 0.4$a$. The theoretical trends of the spectral positions of the three higher order modes, O, H and Q, with respect to $w$ and $R$ are shown in Fig.4a.

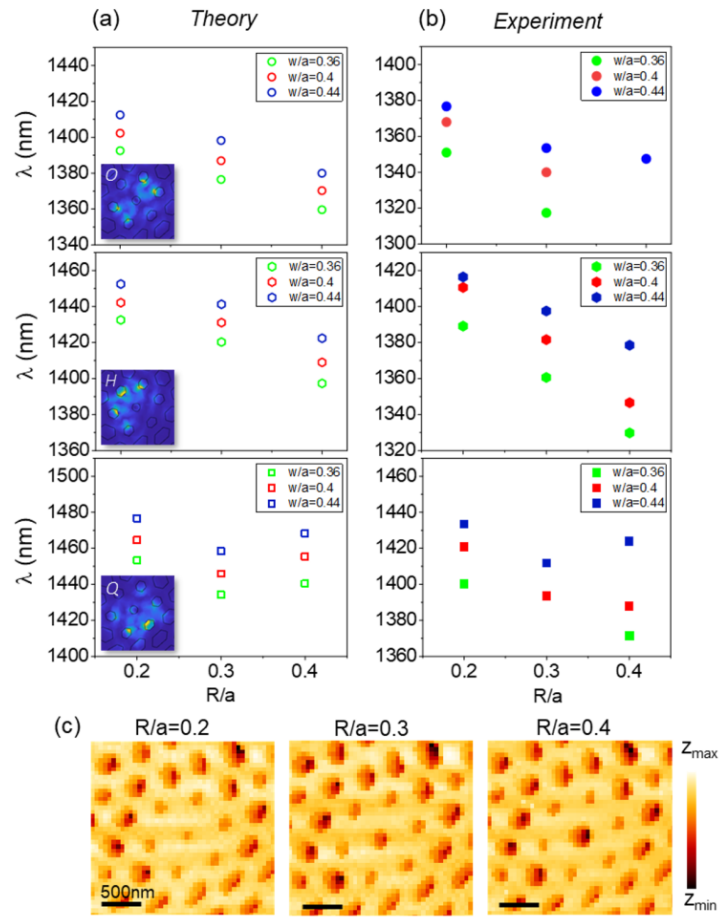

**Figure 4**: **(a)** 3D FEM simulated values and **(b)** SNOM fitted values of the spectral positions of the O, H and Q modes (first, second and third row, respectively) with respect to the central hole radius and different the wall thicknesses $w$=0.36$a$, 0.4$a$ and 0.44$a$. For these structures, $GF$=1. The insets show the FEM maps of the electric field intensity of each mode. **(c)** SNOM Topographies of three cavities with different central hole radius.

On the horizontal axis we report the three values of $R$, while different values of $w$ are highlighted in different colors (green for $w$=0.36$a$, red for $w$=0.4$a$ and blue for $w$=0.44$a$). FEM maps of the electric field intensity are reported in the insets of the graphs. The simulations show an analogous behavior between the studied systems and analogous photonic crystal cavities; indeed, we note an overall trend in two directions, depending on the increase of $R$ and on the increase of $w$: as the central hole radius increases, an "air defect" is formed [36] and the central wavelength of all modes decreases. In contrast, as the wall thickness is increased, we transition towards a "dielectric defect" and the spectral positions of the cavity modes are redshifted. Interestingly, while the O and H modes exhibit shifts that are comparable and approximately constant, the Q mode results more affected by the structural changes and displays non-monotonous shifts with the radius dimension (see the results for $R = 0.4a$), for which an unexpected redshift is observed.

Next, we compare the theoretical results with the SNOM data, reported in Fig.4b. We have performed a scan over the nine cavities with different parameters and extracted the changes in the modes spectral position from Lorentzian fits. Three different topographies for the samples with $R=0.2a$, $0.3a$ and $0.4a$ ($w=0.36a$) are shown in Fig.4c, where the increasing of the air hole at the center of the cavity can be appreciated. The blueshift with the increase of $R$ and the redshift with the increase of $w$ previously found in simulations are faithfully reproduced here. We observe a good agreement between theory and experiment, with the SNOM data reproducing the trends reported in Fig. 4a. However, the O resonances for $w = 0.36a$ and $w = 0.4a$ are not found. It is also possible to observe a $\approx$ 20 nm shift between simulations and experiments. Such deviations between theory and experiment are typical of cavities even in photonic crystal systems and they can be attributed to several uncertainties regarding the difference between nominal and fabricated structural parameters (slab thickness, wall thickness, central hole radius etc).

We now analyze the optical quality factors, $Q_f$ of the localized modes identified. In Fig.5a we report the theoretical $Q_f$ trends for the O, H and Q modes obtained with the same FEM simulations of Fig.4a (we adopt the same color and symbol convention).

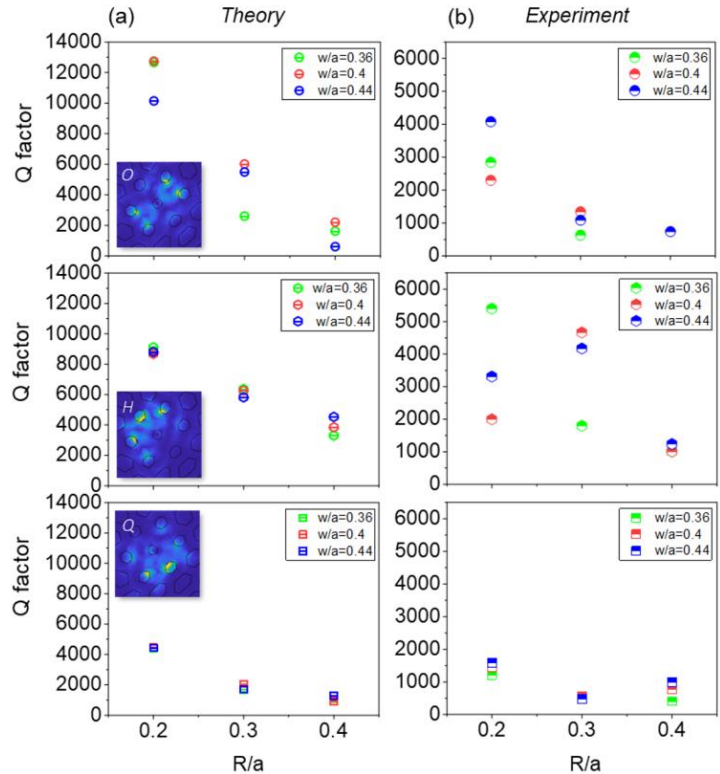

Figure 5: (a) 3D FEM simulated values and SNOM fitted values (b) of the modes $Q_f$ for modes O, H and Q (first, second and third row, respectively full (empty) half-circles, half-hexagons and half-squares) with respect to the values of the central hole radius and different values of the wall thickness $w=0.36a$, $0.4a$ and $0.44a$ (respectively in green, red and blue). The insets show the FEM maps of the electric field intensity of each mode. For these structures, $GF=1$.

The values of $Q_f$ in FEM simulations are calculated through non-hermitian perturbation theory [37]. The behavior of the $Q_f$ as a function of the structural parameters $R$ and $w$ is less intuitive compared to the spectral position of the modes. The simulation shows a common trend, for all the three modes, with the radius of the cavity hole: by increasing $R$, $Q_f$ decreases. This tendency can be explained by considering that for photonic cavities in a slab configuration, the out of plane losses play an essential important role in determining the overall quality factor; in order to prevent mode leakage in the vertical direction one needs to rely on index confinement [19,36]. As such, smaller values of $R$ translate into a higher effective index for the slab and hence better vertical confinement. On the contrary, the influence of $w$ on the $Q_f$ values seems not to indicate an universal trend. This is most likely due to the fact that the thickness of the dielectric walls in the network affects both the modes associated with the HuD surround and the cavity modes. Either way, $w$ has much less impact on the $Q_f$ values than $R$, especially for H and Q modes. By analyzing the experimental $Q_f$s, extracted from the lorentzian fits of the SNOM measurements and reported in Fig. 5b, we note that generally the highest $Q_f$ values are obtained for $R = 0.2a$. However, experimental $Q_f$ values show a reduction compared to simulated $Q_f$ values, and, interestingly, the reduction is larger for the O modes than for the Q modes. Moreover, while the experimental overall maximum value obtained is 5600 for mode H in the configuration of $R = 0.2a$ and $w=0.36a$, FEM simulations establish that mode O has the highest $Q_f$.

These discrepancies between theory and experiment can be attributed to unavoidable fabrication induced disorder and tip perturbation effects [38,39], but also to the presence of the QDs. Specifically, QDs can be considered also as a channel of loss; this means that the results in terms of losses that we obtain for modes at wavelengths overlapping with the QDs emission bands can be affected by this. In the specific case of Fig.5b, mode O is the closest one [1320-1380] nm to the first QDs band (see graph of Fig.2b), and this might explain why we are not detecting the maximum $Q_f$ as in simulations. Clearly, going from a structure with $GF = 1$ to $GF = 0.9$ a rigid blueshift of almost 100 nm in the spectra is obtained, and for this reason the modes shown in Fig.2 are not ideal candidate for $Q_f$ estimations, despite their high signal to noise ratio. Among the reasons explaining the deviations from theory, one can consider also the roughness of the membrane which is not atomically flat due to the QDs density. The theoretical simulations reported in [19] had predicted engineered modes with $Q_f$ as high as 20000; however, in this work, we deal with samples of finite size surrounded by dielectric unpatterned regions. The FEM results reported in Fig. 5, simulating the fabricated sample, confirm that a maximum $Q_f$ of 10000 should still be expected. In addition, the presence of the SNOM tip perturbation effects and of the QDs further lowers the value of the measured $Q_f$.

In conclusion, we investigated experimentally the near-field properties of localized modes in an engineered optical cavity embedded in a hyperuniform disordered environment. We showed interesting features like the coexistence of Anderson localized modes at the PBG edges with cavity modes exhibiting an experimental quality factor as high as 6000. This is very relevant for practical applications, since the ability of optically characterizing localized modes of different symmetry and frequency in the same physical cavity and to detect light modes with different localization properties can have a great impact on all-optical switching, implementations of linear-optical quantum information processors and single photon sources. Following the insight provided by modified L3 cavities, this work could pave the way for realizing photonic cavities in HuD structures with even higher Q factors, for example by employing more than one engineered hole [1,20]. Moreover, the results presented here can open novel approaches for finely tuning the features of resonant cavity modes that, with respect to their periodic counterparts, are located in a more adaptable and flexible architecture, induced by the isotropy of HuD systems.

REFERENCES
1. 1. Y. Akahane, T. Asano, B. S. Song, and S. Noda, Nature 425 944, (2003).
2. H. Y. Ryu, M. Notomi, and E. Kuramoti, Appl. Phys. Lett. 84, 1067 (2004).
3. W. D. Zhou, J. Sabarinathan, P. Bhattacharya, B. Kochman, E. W. Berg, P. C. Yu, and S. W. Pang, IEEE J. Quantum Electron. 37, 1153 (2001).
4. H. G. Park, S. H. Kim, S. H. Kwon, Y. G. Ju, J. K. Yang, J. H. Baek, S. B. Kim, and Y. H. Lee, Science 305, 1444 (2004).
5. J. C. Knight, Nature, 424, pages847–851 (2003).
6. Y. S. Chan, C. T. Chan, and Z. Y. Liu, Phys. Rev. Lett. 80, 956 (1998).
7. K. Nozaki and T. Baba, Appl. Phys. Lett. 84, 4875 (2004).
8. J. Chaloupka, J. Zarbakhsh, and K. Hingerl, Phys. Rev. B 72, 085122, (2005).
9. M. Florescu, P. J. Steinhardt, and S. Torquato, Phys. Rev. B 80, 155112 (2009).
10. M. Florescu, S. Torquato, and P. J. Steinhardt, PNAS, 106 (49) 20658-20663, (2009).

11. M. C. Rechtsman, H. Jeong, P. M. Chaikin, S. Torquato, and P. J. Steinhardt, Phys. Rev. Lett., 101, 073902, (2008).
12. M. Florescu, S. Torquato, and P. J. Steinhardt, Appl. Phys. Lett. 97, 201103 (2010).
13. M. E. Zoorob, M. D. B. Charlton, G. J. Parker, J. J. Baumberg andM. C. Nett, Nature, 404, pages740–743 (2000).
14. S. K. Kim, J. H. Lee, S. H. Kim, I. K. Hwang, and Y. H. Lee, Appl. Phys. Lett., 86, 031101 (2005).
15. P. T. Lee, T. W. Lu, F. M. Tsai, and T. C. Lu, Appl. Phys. Lett., 89, 231111 (2006).
16. S. Torquato and F.H. Stillinger. Physical Review E, 68, 041113, (2003).
17. M. Florescu, S. Torquato, and P. J. Steinhardt, Phys. Rev. B 87, 165116 (2013).
18. M. Florescu, P. J. Steinhardt, and S. Torquato, Phys. Rev. B 80,155112 (2009).
19. T. Amoah and M. Florescu. Physical Review B, 91, 020201(R), (2015).
20. M. M. Milosevic, W. Man, G. Nahal, P. J. Steinhardt, S. Torquato, P. M. Chaikin, T. Amoah, B. Yu, R. A. Mullen, and M. Florescu, Sci. Rep. 9, 20338 (2019).
21. J. Haberko, L. S. Froufe-Pérez, and F. Scheffold, Nat. Commun., 11, 4867, (2020).
22. G.J. Aubry, L. S. Froufe-Pérez, U. Kuhl, O. Legrand, F. Scheffold, and F. Mortessagne , Physical Review Letters, 125, 127402, (2020).
23. P. D. Garcia, R. Sapienza, J. Bertolotti, M. D. Martin, A. Blanco, A. Altube, L. Vina, D.S. Wiersma, and C. Lopez, Physical Review A 78 (2), 023823, (2008).
24. P.D. Garcia, S. Stobbe, I. Sollner, and P. Lodahl, Physical Review Letters 109 (25), 253902, (2012).
25. M. Burresi, V. Radhalakshmi, R. Savo, J. Bertolotti, K. Vynck, and D. S. Wiersma, Physical Review Letters 108 (11), 110604, (2012).
26. N. Granchi, R. Spalding, M. Lodde, M. Petruzzella, F. W. Otten, A. Fiore, F. Intonti, R. Sapienza, M. Florescu, and M. Gurioli, Adv. Opt. Mat., 2102565, (2022).
27. M. Francardi, L. Balet, A. Gerardino, C. Monat, C. Zinoni, L. H. Li, B. Alloing, N. Le Thomas, R. Houdré, and A. Fiore, Phys. Status Solidi C 3, 3693 (2006).
28. R.D. Batten, F.H. Stillinger and S. Torquato. Applied Physics, 104, 033504, (2008).
29. L.S. Froufe-Perez, M. Engelb, J.J. Sáenz and F. Scheffold, PNAS, 114, 9570–9574, (2017).
30. S.G. Johnson and J.D. Joannopoulos. Optics Express, 2001, 8, 173 (2001).
31. F. Intonti, S. Vignolini, F. Riboli, A. Vinattieri, D. S. Wiersma, M. Colocci, L. Balet, C. Monat, C. Zinoni, L. H. Li, R. Houdré, M. Francardi, A. Gerardino, A. Fiore, and M. Gurioli, Phys. Rev. B, B 78, 041401R (2008).
32. A. Femius Koenderink, M. Kafesaki, B. C. Buchler, and V. Sandoghdar, Phys. Rev. Lett. 95, 153904, (2005).
33. J. Dong and D.A. Drabold, Physical Review Letters, 1998, 80, S0031-9007(98)05414-3.
34. S. Imagawa, K. Edagawa, K. Morita, T. Niino, Y. Kagawa, and M. Notomi, Phys.Rev. B, 2010, 82, 115116.
35. T. Yuan, T. Feng and Y. Xu. Optics Express, 2019, 27, 6483-6494.
36. J. D. Joannopoulos, S. G. Johnson, J. N. Winn, and R. D. Meade, "Photonic crystals: molding the flow of light" Princeton Univ. Press, (2008).
37. P. Lalanne,W. Yan, K. Vynck, C. Sauvan, J. P. Hugonin, "Light Interaction with Photonic and Plasmonic Resonances" Laser and Photonics Review, 12, 1–38, (2018).
38. K. G. Cognée, W. Yan, F. La China, D. Balestri, F. Intonti, M. Gurioli, A. F. Koenderink, and P. Lalanne, Optica, Vol. 6, Issue 3, pp. 269-273 (2019).
39. N. Caselli, T. Wu, G. Arregui, N. Granchi, F. Intonti, P. Lalanne, and M. Gurioli, ACS Photonics, 8, 5, 1258–1263, (2021).